\documentstyle[11pt]{article}
\begin{document}
\begin{titlepage}
\vspace{1.5cm}
\title{
{\center \bf A theoretical investigation of the specific heat
of superlattices in a magnetic field}}
\author{
R. M\'elin$^{(1)}$\thanks{e--mail: melin@crtbt.polycnrs--gre.fr}
and F. Fominaya$^{(2)}$\thanks{e--mail: fominaya@labs.polycnrs-gre.fr}
{}\\
$^{(1)}$ International School for Advanced
Studies (SISSA), Via Beirut 2--4, 34014 Trieste, Italy\\
$^{(2)}$ CRTBT--CNRS, 25 Avenue des Martyrs, BP 166X, 38042
Grenoble CEDEX, France
}
\maketitle
\begin{abstract}
\normalsize
We analyze the specific heat variations as
a function of an external magnetic field
of a simple model of superlattice that includes
(i) in--plane ferromagnetic exchange, (ii)
interplane ferromagnetic exchange, (iii)
dipolar interactions, (iv) magnetocristalline
anisotropy. The calculations are carried out at
the spin wave level. The interplay between the existence
of a canting transition and the anisotropy of
the structure generate non trivial behavior for the
spin wave contribution to the
low temperature specific heat as a function
of an external magnetic field when dipolar
interactions and magnetocristalline anisotropy
are taken into account.
\end{abstract}
\end{titlepage}

\renewcommand{\thepage}{\arabic{page}}
\setcounter{page}{1}
\baselineskip=17pt plus 0.2pt minus 0.1pt
\section{Introduction}
Magnetic multilayers made of alternating ferromagnetic films
and non magnetic films (being the typical thickness of these films several
atomic
monolayers) present
interesting properties
in the presence of an external magnetic field, as
for instance the giant magnetoresistance effect
found in \cite{Magneto} and also in other multilayers
\cite{Magneto2}.
Up to now, many different measurements have been performed
\cite{Revuerecente}\cite{Heinrich}, but to our knowledge,
the
specific heat of magnetic multilayers has never been studied
due to the experimental difficulty of measuring a mass as little as a $\mu
g$. Recent
developments in nanocalorimetry allow now to achieve the necessary
resolution for
such a measurement \cite{Denlinger}, \cite{nous}.
The aim of this article is to present a theoretical
investigation of the low temperature specific heat of
bulk multilayers. Two contributions to the specific heat
are present: first, the contribution of paramagnetic layers
and second, a small contribution of spin waves. We analyze
here the variations of the low temperature
spin wave contribution as a function of the strength
of magnetic field applied perpendicular
to the planes or parallel to the planes.
This contribution could
possibly be measured via a high accuracy specific
heat measurement.
Our goal is not to obtain quantitative accurate results,
but rather to give a scenario for the behavior of the
spin wave excitations and the consequences for the
spin wave contribution to the specific heat.

We thus
work with a simple model including
(i) in--plane ferromagnetic exchange of the order of
$J_{\parallel} \simeq 100$ K. (ii)
interplane antiferromagnetic exchange
of the order of $J_{\perp} \simeq 1$ K.
(iii) in--plane
dipolar interactions with a strength $J_d$
of the order $J_d \simeq 1$ K.
(iv) magnetocristalline anisotropy of the
order $K \simeq 0.5$ K.
The magnetic moments are taken to be localized. This
is only an approximation, but widely used in the
literature (see for instance \cite{revue}).
For simplicity, we study the magnetic multilayers
as a bulk system, and we
use periodic boundary conditions in each direction.
In particular, we do not take into account the
surface spin wave modes (see reference \cite{revue}
and references therein).
Even at this level of approximation,
we find a non trivial behavior for the spin wave part of
the specific heat, associated to the transition from a
canted spin configuration to an aligned spin configuration
as the strength of the magnetic field increases.
This transition was already noticed (see \cite{revue}
and references therein). We also find that, when dipolar
interactions and anisotropy are taken into account,
the anisotropy of the structure implies an
anisotropy of the specific heat with respect
to the magnetic field direction. Namely,
the specific heat with a perpendicular magnetic field
is larger than the specific heat with an in--plane
magnetic field.

\section{Description of the model}
Let us now describe the model and fix the notations.
The $x$ and $y$
axis are chosen parallel to the layers and the $z$ axis
is perpendicular to the layers. Since we are interested in
bulk properties, we use periodic boundary conditions in each
direction. Moreover, the spins lye on a cubic lattice.
The Hamiltonian reads
\begin{equation}
\label{Hamil}
H = - \frac{1}{2} \sum_{{\bf x},{\bf x}'}
{\bf S}_{\bf x}.\Lambda({\bf x},{\bf x}').
{\bf S}_{{\bf x}'}
 - {\bf h} . \sum_{{\bf x}} {\bf S}_{\bf x}
,
\end{equation}
where the exchange tensor ${\Lambda}$ is
\begin{eqnarray}
\Lambda_{\alpha,\beta}({\bf x},{\bf x}') &=&
\label{H1}
J_{\parallel} \delta_{\alpha,\beta}
\left(\delta_{{\bf x},{\bf x}'+{\bf e_x}}+
\delta_{{\bf x},{\bf x}'-{\bf e_x}}
+ \delta_{{\bf x},{\bf x}'+{\bf e_y}} +
\delta_{{\bf x},{\bf x}'-{\bf e_y}} \right)\\
&&
\label{H2}
 -  J_{\perp} \delta_{\alpha,\beta}
\left( \delta_{{\bf x},{\bf x}'+{\bf e_z}}+
\delta_{{\bf x},{\bf x}'-{\bf e_z}} \right)
- J_d d_{\alpha,\beta} ({\bf x},{\bf x}') \delta_{z,z'}\\
&& + K d_{\alpha,\beta}({\bf x},{\bf x'})
\left(\delta_{{\bf x},{\bf x}'+{\bf e_x}}+
\delta_{{\bf x},{\bf x}'-{\bf e_x}}
+ \delta_{{\bf x},{\bf x}'+{\bf e_y}}
+ \delta_{{\bf x},{\bf x}'-{\bf e_y}} \right) \delta_{z,z'}
\label{pseudodip}
,
\end{eqnarray}
the dipolar tensor being
\begin{equation}
d_{\alpha,\beta}({\bf x},{\bf x}') =
\frac{1}{|{\bf x}'-{\bf x}|^{3}} \left(\delta_{\alpha,\beta}
- 3 \hat{\bf u}_{\alpha}({\bf x},{\bf x}').
\hat{\bf u}_{\beta}({\bf x},{\bf x}')
\right)
.
\end{equation}
${\bf e_x}$,${\bf e_y}$,${\bf e_z}$
are the unit vectors in the $x$,
$y$ and $z$ directions, and $\hat{\bf u}({\bf x},{\bf x}')$
is the unit vector in the direction ${\bf x}'-{\bf x}$.
In equation (\ref{Hamil}),
${\bf h}$ is an external magnetic field.
In order to make contact with the literature, one should
remember that $h$ is related to the "true"
magnetic field $H$ by $h = M H$, with $M$
the net magnetic moment.
As far as dipolar interactions are concerned, we take into
account only in--plane dipolar interactions. This is 
justified by the structure of magnetic multilayers:
the ferromagnetic layers are separated by paramagnetic layers,
so that it is reasonable, in a first approximation, to
assume that dipolar interactions are long ranged inside a layer
and that two spins in a different layer do not experience
dipolar interactions.

The term (\ref{pseudodip}) is a magnetocristalline
anisotropy term,
modeled via a short range pseudo dipolar interaction,
as proposed in \cite{Bruno}. We choose the anisotropy
to be small enough so that in a zero magnetic field
the magnetization is in the planes.
This amounts to consider only the
case of magnetic multilayers for which the
thickness of the ferromagnetic
layers is large enough \cite{remarque},
typically larger than $20 \AA$.

\section{Spin wave spectrum in a zero magnetic field without dipolar
interactions and anisotropy}
\label{HP}
This section is devoted to the zero magnetic
field case. We determine the spin wave spectrum
in a zero magnetic field and show that, due
to the three dimensionality of the system, no divergences
appear at the spin wave level (unlike the one dimensional
Heisenberg chain case).

\subsection{Spin wave spectrum}
We first analyze the zero magnetic field spin wave spectrum
and the specific heat of the "minimal" model including only
the ferromagnetic $J_{\parallel}$ and antiferromagnetic
$J_{\perp}$ exchange constants. This model in a zero magnetic
field can be solved at the spin wave level
using the Holstein--Primakov bosons method
\cite{HP}. Introducing two types of bosons $a$ and $b$ for
each type of layer, keeping as usually only the $1/s$ leading
quantum corrections in the semi classical (large $s$)
approach and Fourier transforming the Hamiltonian leads to
\begin{equation}
H = \sum_{\bf q} A_{\bf q} \left( a_{\bf q}^{+} a_{\bf q}
+ b_{\bf q}^{+} b_{\bf q} \right)
+ B_{\bf q} \left( a_{\bf q} b_{- {\bf q}}
+ a_{\bf q}^{+} b_{-{\bf q}}^{+} \right)
,
\end{equation}
with
\begin{eqnarray}
A_{\bf q} &=& J_{\parallel}
\left( 2 - \cos{q_x} - \cos{q_y} \right)
+ J_{\perp}\\
B_{\bf q} &=& J_{\perp} \cos{q_z}
.
\end{eqnarray}
The Hamiltonian is readily diagonalized via the
following Bogoliubov transformation
\begin{eqnarray}
c_{\bf q} &=& \cosh{\varphi_{\bf q}} a_{\bf q}
 - \sinh{\varphi_{\bf q}} b_{-{\bf q}}^{+}\\
d_{\bf q} &=& \cosh{\varphi_{\bf q}} b_{\bf q}
- \sinh{\varphi_{\bf q}} a_{\bf q}^{+}
,
\end{eqnarray}
with $\tanh{(2 \varphi_{\bf q})} = - B_{\bf q}/A_{\bf q}$,
leading to the following spin--wave spectrum
$\omega_{\bf q}^{2} = A_{\bf q}^{2}- B_{\bf q}^{2}$:
\begin{equation}
\label{spectrumh0}
\omega_{\bf q}^{2} = \left(
J_{\parallel}(2 - \cos{q_x} - \cos{q_y}) + J_{\perp}
\right)^{2} - J_{\perp}^{2} \cos^{2}{q_z}
.
\end{equation}
If $J_{\parallel}=0$, we recover the spin--wave spectrum
$\omega_{\bf q} = J_{\perp} | \sin{q_z} |$ of the
Heisenberg chain \cite{Affleck}
\cite{Kittel} and if $J_{\perp}=0$, we recover the
$\omega_{\bf q} = J_{\parallel}(2-\cos{q_x}
- \cos{q_y})$ spectrum of the
two--dimensional ferromagnet \cite{Kittel}.

\subsection{Stability of the classical alternate ground state}
Before going further, we would like to show that the
antiferromagnetic perpendicular exchange $J_{\perp}$
does not destabilize the alternate ground state,
unlike the one--dimensional case where the N\'eel state
is destabilized by the first quantum corrections.
The stability in the multilayer case is due to the
three dimensionality of the structure. We calculate
the first quantum corrections to the magnetization.
To do so, we notice that
\begin{equation}
\langle a_{\bf q}^{+} a_{\bf q} \rangle =
\frac{1}{2} \left( \frac{1}{\sqrt{1 - \tanh^{2}{(
2 \varphi_{\bf q}})}} -1 \right)
,
\end{equation}
leading to the average quantum correction
$\Delta M$ to the magnetization per spin
in a layer:
\begin{equation}
\Delta M = - \frac{1}{(2 \pi)^{3}}
\int_{BZ} d {\bf q} \left( \frac{1}{\sqrt{1 - \tanh^{2}
{(2 \varphi_{\bf q})}}}-1 \right)
,
\end{equation}
where BZ denotes an integral over the
first Brillouin zone.
This integral is infrared convergent:
\begin{equation}
\Delta M \sim - \frac{1}{(2 \pi)^{3}}
\int_{BZ} d {\bf q} \frac{1}{\sqrt{q_z^{2} + \frac{J_{\parallel}}
{J_{\perp}}(q_x^{2} + q_y^{2})}}
,
\end{equation}
where we have expanded the $A_{\bf q}$ and $B_{\bf q}$
in the small ${\bf q}$ limit. A dimensional analysis
shows that $\Delta M \sim J_{\perp}/J_{\parallel}$.
In the limit $J_{\parallel} \rightarrow 0$, we recover
the spin--wave divergence of the Heisenberg chain \cite{Affleck}.
In the multilayers, $J_{\parallel}$ is two orders of
magnitude larger than $J_{\perp}$: our spin wave approach
is thus consistent from the point of view of the absence
of divergences of the spin wave theory. On the other hand,
the three--dimensionality of the structure also avoids
problems related to the Mermin--Wagner theorem in two
dimensions at a finite temperature (absence of spontaneous
order in two dimension for continuous spins). We have
shown the consistency of our spin wave approach only in the
case of a zero magnetic field and without dipolar interactions.
However, the spin wave approach is still consistent in
the presence of a magnetic field and with dipolar interactions.

\section{Spin wave theory in a perpendicular magnetic field}
\subsection{Classical ground state without dipolar interactions
and anisotropy}
\label{classical}
We are first going to examine the spin wave theory
when an external magnetic field is switched on, but
in the absence of dipolar interaction and
anisotropy. The effect of
dipolar and anisotropy interactions will be discussed
later on.
At the classical level, the zero temperature ground state
corresponds to the condition that each spin is exactly aligned
in its local field. Because of the translational invariance,
each spin in a given layer points in the same direction
(at the classical level and at zero temperature).
We denote by ${\bf n}^{(a)}$ the spin
vector in layers of type (a),
and ${\bf n}^{(b)}$ the spin vector
in (b) type layers: the spins can be canted
in a layer by layer fashion.
Layers (a) and (b) are alternating
along the $z$ direction.
The local field ${\bf h}^{(a),(b)}$ in layers
(a) and (b) respectively is given by
\begin{eqnarray}
{\bf h}^{(a)} &=& {\bf h}
+ 2 J_{\parallel} {\bf n}^{(a)} -
J_{\perp} {\bf n}^{(b)}\\
{\bf h}^{(b)} &=& {\bf h}
+ 2 J_{\parallel} {\bf n}^{(b)} -
J_{\perp} {\bf n}^{(b)}
,
\end{eqnarray}
where ${\bf h}$ is the external magnetic field.
We denote by ${\bf u}$ the unit
vector in the direction of the external magnetic field:
${\bf h} = |h| {\bf u}$.
We are looking for a classical ground state
such as ${\bf n}^{(a)}$, ${\bf n}^{(b)}$ and ${\bf u}$
are coplanar. As we will see below, such a classical
spin configuration exists. We denote by $\alpha$
the angle between the vectors ${\bf u}$ and ${\bf n}^{(a)}$:
${\bf n}^{(a)} = {\bf u} \cos{\alpha} +
{\bf v} \sin{\alpha}$,
where ${\bf v}$ is a vector orthogonal to ${\bf u}$.
The angle $\alpha$ between the magnetic field
and the spin direction is called the canting angle
in what follows.
The local field ${\bf h}^{(a)}$ experienced by a
spin in a type (a) layer is easily expressed as
\begin{equation}
{\bf h}^{(a)} = {\bf u} \left(
h + (2 J_{\parallel} - J_{\perp}) \cos{\alpha} \right)
+ {\bf v}
\left( 2 J_{\parallel} + J_{\perp} \right) \sin{\alpha}
= |{\bf h}^{(a)}|  \left( {\bf u} \cos{\alpha}
+ {\bf v} \sin{\alpha}  \right)
.
\end{equation}
The last equality simply means that, at the classical
level, the spins are aligned with their local field.
We can thus determine the direction of the ${\bf n}^{(a)}$
vector:
\begin{equation}
\label{condition}
\begin{array} {cl}
\cos{\alpha} = h/2 J_{\perp} & \mbox{if
$h \le 2 J_{\perp}$}\\
\alpha = 0 & \mbox{if $h
\ge 2 J_{\perp}$}
.
\end{array}
\end{equation}
At the classical level,
the spins are thus exactly aligned with the
magnetic field provided the intensity of the magnetic field
is larger than $2 J_{\perp}$ and in the
absence of dipolar interactions and anisotropy. 
\subsection{Spin waves spectrum without dipolar interactions
and anisotropy}
\label{isotrop}
In the absence of dipolar interactions and anisotropy,
the classical canted configuration of spins is given
by (\ref{condition}). The in--plane spin wave degrees of
freedom decouple from the perpendicular degrees of
freedom. The contribution of the in--plane spin wave
degrees of freedom is a trivial ferromagnetic term,
which will be added at the end of the calculation.
For the moment we consider only spin wave excitations
along the $z$ direction, that is, we are going to
solve the Heisenberg chain with an antiferromagnetic
exchange $J_{\perp}$
in a magnetic field
at the spin wave level. In order to obtain the appropriate
excitations, we first need to express the spin operators
$\hat{\sigma}^{\pm,z}$ in terms of the rotated spin
operators $\hat{\sigma}^{\pm,z}_R$ corresponding to 
a spin quantization along the direction of the classical
ground state. The rotated spin states are
\begin{eqnarray}
|+\rangle_R &=& \cos{\left( \frac{\alpha}{2} \right)}
|+ \rangle -i \sin{\left( \frac{\alpha}{2} \right)}
|- \rangle\\
|-\rangle_R &=& \cos{\left( \frac{\alpha}{2} \right)}
|- \rangle -i \sin{\left( \frac{\alpha}{2} \right)}
|+ \rangle
,
\end{eqnarray}
where the rotated spin states $|\pm\rangle_R$
are deduced from the $|\pm \rangle$ states by a rotation
of angle $\alpha$ around the $x$ axis.
This allows us to express
the spin operators in the laboratory framework
in terms of the spin operators in the local
framework:
\begin{eqnarray}
\hat{\sigma}^{+} &=& - \frac{i}{2} \sin{\alpha}
\hat{\sigma}^{z}_R + \cos^{2}{(\alpha/2)}
\hat{\sigma}^{+}_R + \sin^{2}{(\alpha/2)}
\hat{\sigma}^{-}_R\\
\hat{\sigma}^{-} &=& \frac{i}{2} \sin{\alpha}
\hat{\sigma}^{z}_R + \cos^{2}{(\alpha/2)}
\hat{\sigma}^{-}_R + \sin^{2}{(\alpha/2)}
\hat{\sigma}^{+}_R\\
\hat{\sigma}^{z} &=& \cos{\alpha} \hat{\sigma}^{z}_R
-i \sin{\alpha} \hat{\sigma}^{+}_R + i \sin{\alpha}
\hat{\sigma}^{-}_R
.
\end{eqnarray}
The expressions should be plugged into
the Hamiltonian of the Heisenberg
chain in a magnetic field. The spin wave approximation
then consists in using boson operators
related to the local spin operators as follows:
$\hat{\sigma}^{+}_R=a$, $\hat{\sigma}^{-}_R=a^{+}$
and $\hat{\sigma}^{z}_R=1-2 a^{+}a$.
Technically, the effect of the rotation
to the local framework is to eliminate
linear terms in boson operators. We are
thus left with a hamiltonian that is
quadratic in terms of Holstein--Primakov
bosons.
Considering only two consecutive spins
with boson operators $a$ and $b$, and after
straightforward calculations, we obtain
the following spin wave Hamiltonian
\begin{equation}
\label{HSW}
H^{(sw)}=\frac{J_{\perp}}{2} \left(a^{+}a
+b^{+}b \right) + \frac{J_{\perp}}{2} \sin^{2}{\alpha}
\left(a^{+}b^{+}+ab \right) + \frac{J_{\perp}}{2}
\cos^{2}{\alpha}\left(ab^{+}+a^{+}b\right)
.
\end{equation}
After Fourier transforming the chain Hamiltonian
and performing a Bogoliubov rotation, we deduce
the following spin wave spectrum
$\omega_{q_z}^{2}=A_{q_z}^{2}-B_{q_z}^{2}$ with
\begin{eqnarray}
\label{A1}
A_{q_z}^{(\perp)}&=&J_{\perp}\left(1+\cos^{2}{\alpha}\cos{q_z}\right)\\
B_{q_z}^{(\perp)} &=& J_{\perp} \sin^{2}{\alpha} \cos{q_z}
.
\label{B1}
\end{eqnarray}
This expression is defined on the unfolded
Brillouin zone $[-\pi,\pi]$. However, one should
remember that, because of the doubling of the unit
cell, the ``true'' Brillouin zone is $[-\pi/2,\pi/2]$.
In order to come back from the unfolded
Brillouin zone $[-\pi,\pi]$ to the ``true'' Brillouin zone
$[-\pi/2,\pi/2]$,
one should consider two spin wave modes for a
given wave vector. Notice also that the unfolded
dispersion relation is not unique: starting from
two distinct spin wave modes in the ``true''
Brillouin zone, there are two ways to unfold
the dispersion relation to the $[-\pi,\pi]$
interval. The spectrum defined by (\ref{A1})
and (\ref{B1}) corresponds only to one
of the unfolding procedures. The other unfolding
procedure corresponds to changing $q_z$
into $q_z+\pi$ in the unfolded
Brillouin zone.

We notice that, in a zero
magnetic field, there are two Goldstone modes.
As the magnetic field is switched on, one of these
Goldstone modes acquires a gap while the other one
remains massless (rotations around the magnetic field).
The dispersion relation defined by (\ref{A1}) and (\ref{B1})
is thus consistent with the fact that, at the classical level
in the presence of a non zero magnetic field the classical
ground state is defined up to a rotation around
the magnetic field, thus leading a single gapless mode
at the spin wave level.

We have rederived the spin wave spectrum defined
by (\ref{A1}) and (\ref{B1}) using a method
based on the semiclassical equations of motion,
which lead exactly to the same result. We do not
present here these calculations since they do not
lead to new physical insights.

In order to come back to the multilayer problem, we
have to take into account the existence of
in--plane ferromagnetism, leading to an additional
$A$ term that should be added to (\ref{A1}):
\begin{equation}
A_{q}^{(\parallel)}=J_{\parallel}\left(2-\cos{q_x}-\cos{q_y}\right)
.
\end{equation}

Now, if $h>2 J_{\perp}$, the spins are aligned with
the magnetic field. The dispersion relation is
straigthforwardly found to be
\begin{equation}
\omega_q = h-J_{\perp} + J_{\perp} \cos{q_z}
+ J_{\parallel}
\left(2-\cos{q_x}-\cos{q_y} \right)
.
\end{equation}
This leads to a gap $\Delta=h-2 J_{\perp}$.

We have thus obtained an analytic expression
for the spin wave spectrum in the absence of
dipolar interactions and anisotropy.
The evolution of the dispersion relation
for $q_x=q_y=0$ as a function of $q_z$ and
the magnetic field is plotted on figure
\ref{Fig1}.

We emphasize that this result does not depend on the
direction of the magnetic field: since the Hamiltonian
without magnetic field is spin rotational invariant,
the spin wave spectrum should be the same whatever the
direction of the magnetic field.

\subsection{Specific heat without dipolar interactions and
anisotropy}
The specific heat per spin is obtained from the
dispersion relation via
\begin{equation}
c_v = \frac{1}{4 T^{2}} \int d {\bf q}
\frac{\omega_{\bf q}^{2}}{\sinh^{2}{\left(\beta \omega_{\bf q}
/2\right)}}
,
\end{equation}
where $\beta=1/T$ is the inverse temperature.
We calculated numerically the spin wave specific heat
as a function of a magnetic field for a low temperature
$T=3 K$. The result is plotted on the curve (a)
of figure \ref{Fig2}. We observe the existence
of two regimes: (i) $h<h_c=2 J_{\perp}$: the
specific heat is slowly decaying as the magnetic
field is increased. We attribute this slow decay
to the fact that the system remains gapless
in the presence of the magnetic field. If $h=0$,
there are two Goldstone modes. One of these Goldstone
modes becomes massive as the magnetic increases,
leading to a decrease of the low energy
density of states and thus a slow decrease in
the specific heat. (ii) $h>h_c$: the system
acquires a gap $\Delta=h-2 J_{\perp}$,
leading to a decay to zero of the specific heat
as the magnetic field increases, with a cross--over
magnetic field $h^{*}=2 J_{\perp} + T$.

\subsection{Spin wave spectrum in the presence of
in--plane dipolar interactions}
We now assume the existence of in--plane dipolar interactions.
This assumption is motivated by the fact that the
ferromagnetic layers are separated by paramagnetic layers,
and it is thus reasonable to suppose that only spins in the
same layer are coupled by dipolar interactions.

The first step is to calculate the zero temperature
classical configuration of spins. This can be easily done
by minimizing the classical energy with respect to the
canting angle and we find
\begin{equation}
\label{canteddip}
\cos{\alpha} = \frac{h}{2 J_{\perp} + 3 \pi J_d}
\end{equation}
in the regime where the spins are canted. Physically,
we expect that, at the classical level, the effect of
dipolar interactions is to open the canted angle
since dipolar interactions play in favor of
attracting the spins in the planes. This is indeed
the case, as shown by equation (\ref{canteddip}).
Notice that in the evaluation of the dipolar
interaction energy, we have replaced the
discrete summation over the lattice by an integral.
We will use this approximation throughout this paper
to treat the effects of long range in--plane dipolar
interactions.

Next, we need to calculate the spin wave spectrum
in the presence of dipolar interactions. The calculations
are lengthly but straightforward so that we just give
here the results. The idea consists in eliminating
linear Holstein--Primakov bosons by a rotation
to the local framework. Keeping only the quadratic
terms in boson operators leads to the following results
in the canted phase. The additional contribution to
the $A_{q}$ term due to dipolar interactions
is
\begin{equation}
\label{Adip}
A_q^{(dip)} =\frac{3 \pi J_d}{2} \cos^{2}{\alpha}
+\pi J_d \left(I_0(q_{\parallel})-1 \right)
+ \frac{3 \pi J_d}{4} I_0(q_{\parallel}) \left(
3 \sin^{2}{\alpha}-2 \right)
+\frac{9 \pi J_d}{4} \sin^{2}{\alpha}
\cos{(2 \phi_q}) I_2(q_{\parallel})
,
\end{equation}
where $\phi_q$ is the angle between the
vector $(q_x,q_y)$ and the $x$ axis, and
$I_n$ is expressed in terms of Bessel functions as
\begin{equation}
I_n(q) = \int_1^{+ \infty} \frac{dr}{r^{2}}
J_n(q r)
.
\end{equation}
The dipolar contribution to the $B_q$ terms is
\begin{equation}
\label{Bdip}
B_q^{(dip)} = - \frac{3 \pi J_d}{4}
\left( \sin^{2}{\alpha} I_0(q_{\parallel})
- \cos{(2 \phi_{q})}I_2(q_{\parallel})
\left((1+\cos^{2}{\alpha}) - 2 i \cos{\alpha}
\right) \right)
.
\end{equation} 
As expected, if $q_x=q_y=0$, $|A_0^{(dip)}|=
|B_0^{(dip)}|$ since the $q_x=q_y=0$ mode
in a single ferromagnetic layer with dipolar
interactions is gapless.
Notice that the contribution (\ref{Bdip}) is in general
a complex number. However, this is not a problem:
using a phase transformation on the Holstein--Primakov
bosons, we show that the spin wave spectrum is
given by $\omega_q^{2}=(A^{tot}_q)^{2}-|B^{tot}_q|^{2}$,
where the subscript ``tot'' refers to the total
$A$ and $B$ terms (the contributions of
ferromagnetic in--plane exchange terms
+ antiferromagnetic coupling terms + dipolar interaction
terms have been added). Specializing the dispersion relation
to the $q_x=q_y=0$ case, and in the absence of a
magnetic field ($\alpha=\pi/2$), we observe the
existence of two Goldstone modes in the absence of
dipolar interactions but in the presence of
dipolar interactions, only one of these modes
remains massless whereas the other one acquires
a gap. This remaining massless mode corresponds to
rotations of the classical spins inside the planes.
In the presence of a magnetic field, this massless
mode remains since it also corresponds to the
rotation of the spins around the magnetic field
(perpendicular to the planes).

Notice also that the terms proportional
to $I_2(q_{\parallel})$ in (\ref{Adip})
and (\ref{Bdip}) are expected to lead
to a very small (and neglectable) contribution
to the specific heat: the main contribution
to the specific heat comes from the lowest
energy states, namely the states next to
$q_x=q_y=0$. Along this line, the terms
proportional to $I_2$ vanish since $I_2(0)=0$.
In the rest of the Brillouin zone the contribution
of the terms involving $I_2$ remains very small
compared to the in--plane ferromagnetic exchange term,
since $J_{\parallel}$ is two orders of magnitude larger
than $J_d$. In the present case of a perpendicular
magnetic field, the terms proportional to $I_2$
can be treated exactly since they are the same for
all types of layers (invariance under the
$\alpha \rightarrow - \alpha$ transformation).
As we shall see below, this is not the case
for a magnetic field parallel to the layers.
However, we will argue that one can safely neglect these
terms since they lead to a neglectable contribution
to the specific heat.

The dispersion relation on the line $q_x=q_y=0$
as a function of $q_z$ and the magnetic field
is plotted on figure \ref{Fig3}.

\subsection{Magneto cristalline
anisotropy as a pseudo dipolar interaction}
We now take into account an anisotropy term (\ref{pseudodip}),
under the form of a pseudo dipolar in--plane interaction.
We just give the results. One should first calculate
the canting angle in the presence of such an interaction.
After minimizing the classical energy, we find easily
\begin{equation}
\cos{\alpha} = \frac{h}{2 J_{\perp} + 3 \pi J_d/2-3K}
.
\end{equation}
This equation is consistent with the fact that
the pseudo--dipolar interactions play in
favor of aligning the spins perpendicular to the
layers, and thus reduce the canting angle.
As far as the spin wave spectrum is concerned,
the first effect of these pseudo--dipolar interactions
is to renormalize $J_{\parallel}$: $J_{\parallel}
\rightarrow J_{\parallel}-K$. Since $K$ is
much smaller than $J_{\parallel}$, this effect is
very small. The contribution of the anisotropy
to $A$ is found to be
\begin{equation}
A_{q}^{(K)} = - 3 K
+ \frac{3 K}{2} \left( \cos{q_x} + (\cos^{2}{\alpha})
\cos{q_y} \right)
,
\end{equation}
and the contribution to $B$ is
\begin{equation}
B_{q}^{(K)} = \frac{3 K}{2} \left(\cos{q_x} - (\cos^{2}{\alpha})
\cos{q_y} \right)
.
\end{equation}
As in the case of dipolar interactions,
$|A_0^{(K)}|=|B_0^{(K)}|$ if $q_x=q_y=0$ since
the $q_x=q_y=0$ mode is a Goldstone mode
of a ferromagnetic layer with pseudo--dipolar interactions.
Figure \ref{Fig4} shows the dispersion relation
along the line $q_x=q_y=0$ as a function of $q_z$
and the magnetic field. Qualitatively, adding an anisotropy
term does not change the main features of the dispersion
relation (namely the existence of a single massless
Goldstone mode).

\subsection{Specific heat in the presence of
in--plane dipolar interactions and magneto cristalline anisotropy}
We are now in position to calculate the variations of the
low temperature specific heat as a function of the intensity of
a magnetic field applied perpendicular to the layers,
within the spin wave approximation for the Hamiltonian
(\ref{H1})(\ref{H2})(\ref{pseudodip}). These variations
are plotted on the curve (b) of figure \ref{Fig2}.
We observe two regimes: (i) for $h<h_c=2 J_{\perp}
+ 3 \pi J_d - 3 K$, we observe that the specific heat
increases as the magnetic field increases.
We attribute this behavior to the fact that
increasing the magnetic field induces an increase
of the low energy density of states. This can be seen
as follows. If $q_x=q_y=0$, the dispersion
relation around the Goldstone mode $q_z=\pi-\Delta q_z$
(in the unfolded Brillouin zone) is of the
form
\begin{equation}
\omega_{q_x=0,q_y=0,q_z=\pi-\Delta q_z}
= \lambda \sin{\alpha} \Delta q_z
.
\end{equation}
with $\lambda$ independent on the magnetic field:
\begin{equation}
\lambda^{2} = J_{\perp}\left(J_{\perp}
+ \frac{3 \pi J_d}{2} - \frac{3 K}{2} \right)
.
\end{equation}
The density of states in the vicinity of the
massless mode is thus an increasing function of
the magnetic field, leading to an increase of the
specific heat as the magnetic field increases.

(ii) $h>h_c$: the gap is $\Delta=h-h_c$ thus leading
to a decrease to zero of the specific heat as the magnetic 
field increases with a cross--over
magnetic field $h^{*}=T+h_c = T + 2 J_{\perp}
+ 3 \pi J_d/2 - 3 K$.

Finally, we can compare this result with the
specific heat in the absence of dipolar interactions.
In a zero magnetic field, we expect the specific heat
without dipolar interactions and anisotropy
to be larger than the specific heat with dipolar interactions
and anisotropy since there is a single massless mode
in the former case and two massless modes in the latter. This is
indeed the case, as shown on figure \ref{Fig2}.
Now, consider the aligned state ($h>h_c$).
The gap is reduced when
dipolar and pseudodipolar interactions are switched on,
so that switching on dipolar interactions and anisotropy
should increase the specific heat, which is what is observed
on figure \ref{Fig2}. The existence of a crossing point
between the specific heat curves with and without
dipolar interactions and anisotropy on figure \ref{Fig2}
is thus consistent with our qualitative analysis.


\section{Spin wave theory in a longitudinal magnetic field}
We now examine the case of a magnetic field applied parallel
to the layers. The effect of the orientation of the
magnetic field can only be seen when dipolar interactions
and anisotropy are taken into account. Namely, in the
absence of dipolar interactions and anisotropy,
the spin wave spectrum is the one calculated in section
\ref{isotrop}. We are going to calculate
the effects of a longitudinal magnetic field. To do so,
we use the same technique as in section \ref{isotrop}:
we first express the Hamiltonian in terms of the
spin operators quantized along the local magnetization
axis, which leads to a quadratic Hamiltonian
at the spin wave level, that can be diagonalized via
a Fourier transform and a Bogoliubov transformation.
First, we treat the problem in the absence of
dipolar interactions and anisotropy with the
spins in the planes direction. Of course, we recover
the results of section \ref{isotrop}. In a second step,
we discuss the effects of in--plane dipolar interactions
and anisotropy. In order to avoid confusion
with the previous calculations, we call $\varphi$
the canting angle.

\subsection{Spin wave spectrum in the absence of
dipolar interactions and anisotropy}
If ${\bf w}$ denotes the unit vector of
coordinates ${\bf w}(\sin{\varphi},\cos{\varphi},0)$,
with $\varphi$ the canting angle, then,
the local spin states in a given layer are
\begin{eqnarray}
|+\rangle_R &=& \frac{1}{\sqrt{2}}\left(
|+\rangle + e^{-i \varphi}|-\rangle\right)\\
|- \rangle_R &=& \frac{1}{\sqrt{2}} \left(
-e^{i \varphi} |+\rangle + |-\rangle \right)
,
\end{eqnarray}
where we have performed a rotation of $\pi/2$
around the ${\bf w}$ axis.
The case of the other type of layer is
obtained by replacing $\varphi$ by $- \varphi$
in these equations. The expression of the
spin operators $\hat{\sigma}^{\pm,z}$ in terms of
the local spin operators $\hat{\sigma}_R^{\pm,z}$ is
\begin{eqnarray}
\hat{\sigma}^{+} &=& \frac{1}{2} \left(
e^{-i \varphi} \hat{\sigma}^{z}_R + \hat{\sigma}^{+}_R
- e^{-2 i \varphi} \hat{\sigma}^{-}_R \right)\\
\hat{\sigma}^{-} &=& \frac{1}{2} \left(
e^{i \varphi} \hat{\sigma}^{z}_R + \hat{\sigma}^{-}_R
- e^{2 i \varphi} \hat{\sigma}^{+}_R \right)\\
\hat{\sigma}^{z} &=& -e^{i \varphi} \hat{\sigma}^{+}_R
- e^{-i \varphi} \hat{\sigma}^{-}_R
.
\end{eqnarray}
Using the same technique as in section \ref{isotrop},
we express the Hamiltonian of a single a--b link in
terms of the $a$ and $b$ bosons (one type of boson
for each direction of the classical ground state).
In order to get rid of phase factors, one has to
define new boson operators $\tilde{a}=a \exp{(i \varphi)}$
and $\tilde{b} = b \exp{(-i \varphi)}$. We then
recover the spin wave Hamiltonian (\ref{HSW})
in terms of the $\tilde{a}$ and $\tilde{b}$
bosons.

\subsection{Spin wave contribution of in--plane dipolar
interactions}
The first step is to determine the classical spin
configurations. In the presence of dipolar interactions,
the classical spins are aligned in the planes, with
a canting angle
\begin{equation}
\label{CANTED}
\cos{\varphi} = \frac{h}{2 J_{\perp}}
.
\end{equation}
Notice that the dipolar interaction coupling $J_d$
does not come into account in the expression of the
canting angle. This is due to the fact that the
spins are in the planes, and therefore the classical
energy of dipolar interactions is a constant,
independent on the canting angle $\varphi$.
This is already a first difference between the
cases of a perpendicular and longitudinal
magnetic fields.

The second step is to include in--plane dipolar interactions
at the spin wave level. We do not give the details of the
calculations. The contribution of dipolar interactions
to $A$ for a given layer is
\begin{equation}
\label{Adip2}
A^{(dip)}_q = \pi J_d \left( I_0(q_{\parallel})-1 \right)
+ \frac{3 \pi J_d}{4} I_0(q_{\parallel})
- \frac{9 \pi J_d}{4} \cos{(2(\phi_q-\varphi))}
I_2(q_{\parallel})
,
\end{equation}
and the contribution to $B$ is
\begin{equation}
\label{Bdip2}
B^{(dip)}_q = \frac{3 \pi J_d}{4}
\left( I_0(q_{\parallel}) + \cos{(2(\phi_{q}+\varphi))}
I_2(q_{\parallel}) \right)
.
\end{equation}
These calculations have been carried out in a given layer.
The main problem regarding these expressions is that
they are not invariant under the $\varphi \rightarrow
- \varphi$ transformation. In other words, the dipolar
terms induce a hybridization between the two Goldstone
modes in the $[-\pi/2,\pi/2]$ Brillouin zone.
However the strength of the hybridization is
proportional to $I_2(q_{\parallel})$.
As mentioned earlier, we expect
the terms proportional to $I_2$ to lead to a really
small contribution to the specific heat since
i) $I_2$ vanishes along the line $q_x=q_y=0$,
where the lowest energy modes are located
ii) outside the $q_x=q_y=0$ line, these terms
are dominated by the ferromagnetic exchange interaction
which is two orders of magnitude larger than the dipolar
coupling. We can thus safely neglect the terms proportional
to $I_2$ in our specific heat calculation.

Finally, in the presence of dipolar interactions,
the results in a perpendicular and longitudinal magnetic
field should be the same in the limit of vanishing fields.
This is indeed the case. In order to get the
same result for $\alpha=\varphi=\pi/2$, one should
make a $\pi$ shift on the $z$ component of the wave vector.
This $\pi$ shift is allowed since, as mentioned earlier, there
are two possibilities of unfolding the ``true'' Brillouin
zone $[-\pi/2,\pi/2]$ to the extended Brillouin zone
$[-\pi,\pi]$. These two possibilities are precisely
related to each other by a $\pi$ shift on the $z$ component
of the wave vector.

Before discussing the results (\ref{Adip2}) and (\ref{Bdip2}),
we would like to include the effects of an anisotropy.

\subsection{Spin wave contribution of the
magneto cristalline anisotropy
as a pseudo--dipolar interaction}
The magnitude of the anisotropy is chosen to be small
enough so that the classical configuration of spins
is aligned in the plane, whatever the value
of the longitudinal magnetic field. Under this assumption,
the canting angle is given by (\ref{CANTED})
and is independent on the strength $K$ of
the anisotropy. The anisotropy term
can straightforwardly be incorporated
in the spin wave approach. The results are as follows.
The first effect of the anisotropy is
a renormalization of $J_{\parallel}$:
$J_{\parallel} \rightarrow J_{\parallel} - K$.
This effect is very small since $J_{\parallel}$
is two orders of magnitude larger than $K$.
The contribution of the anisotropy to the
$A$ term is
\begin{equation}
A^{(K)} = -3 K + \frac{3 K}{2} \left(
\sin^{2}{\varphi} \cos{q_x}
+ \cos^{2}{\varphi} \cos{q_z} \right)
.
\end{equation}
The contribution of the anisotropy to the
$B$ term is 
\begin{equation}
B^{(K)} = - \frac{3 K}{2} \left( \sin^{2}{\varphi}
\cos{q_x} + \cos^{2}{\varphi} \cos{q_z} \right)
.
\end{equation}
The dispersion relation in the presence of
an anisotropy and dipolar interactions
is plotted on figure \ref{Fig5} in the
extended Brilouin zone for $q_x=q_y=0$.
In a zero magnetic field, we observe
the presence of a massless mode associated
to the rotation of the spins inside the planes.
As soon as a magnetic field is switched on,
this mode acquires a gap, due to the fact that
the magnetic field breaks the rotational invariance
of the ground state with respect to in--plane rotations.
Another effect is that if $h=2 J_{\perp}$, we observe
the existence of a massless mode. The existence of this
gapless mode can be understood at the classical
level as follows: consider a magnetic field strictly
equal to $2 J_{\perp}$. It is then possible to calculate
the energy of a classical spin wave of amplitude
$\theta$ around the aligned state, and expand the
energy in powers of the amplitude $\theta$ of
the classical spin wave. One then get the result
that the leading term in this expansion
is of order $\theta^{4}$. More precisely,
the energy per spin of an antiferromagnetic classical
spin wave of amplitude $\theta$ around the aligned state
($h \ge 2 J_{\perp}$) is
\begin{equation}
E(\theta) = -\frac{h}{2} \cos{\theta} + \frac{J_{\perp}}{4}
\cos{(2 \theta)} = -\frac{1}{4}(h-2 J_{\perp})(1-\theta^{2})
- \frac{1}{48}(h-8 J_{\perp}) \theta^{4} + ...
\end{equation}
If $h=2 J_{\perp}$ the leading term of this expansion
is of order $\theta^{4}$,
thus leading to a flat
mode at the harmonic order.
It is thus
not surprising that at the quantum spin wave level one finds
a gapless mode right at the canting transition.

\subsection{Specific heat in a longitudinal
magnetic field}
The specific heat in a longitudinal magnetic field
is plotted on the curve (c) of figure \ref{Fig2}.
We observe the same type of variations as in the
perpendicular magnetic field case, except that
the specific heat with an in--plane magnetic field
is smaller than the specific heat without dipolar
interactions and anisotropy. This can be explained qualitatively
as follows. In the aligned state ($h>2 J_{\perp}$),
the gap is given by
\begin{equation}
\Delta^{2} = \left(h - 2 J_{\perp} \right)^{2}
+ (h-2 J_{\perp})\left( \frac{3 \pi J_d}{2}-3K \right)
,
\end{equation}
where the dispersion relation has been specialized
to the $q_x=q_y=0$, $q_z=\pi$ case. As a consequence, the
gap with an in--plane magnetic field and dipolar interactions
and anisotropy is larger than in the absence of dipolar interactions
and anisotropy.
We thus expect from this argument a lower specific heat,
which is indeed the case as observed on figure \ref{Fig2}.

\section{Conclusion and discussion}
\label{discussion}
Let us now summarize our results. We have calculated the
spin wave contribution to the specific heat in three
different cases: (a) without dipolar interactions
and anisotropy. At this level of approximation,
the specific heat does not depend on the direction
of the magnetic field since the Hamiltonian without a
magnetic field is spin rotational invariant.
(b) with dipolar interactions and anisotropy for a perpendicular
applied magnetic field.
(c) with dipolar interactions and anisotropy for a longitudinal
applied magnetic field.
In case (b), the location of the canting transition depends
explicitly on the strength of dipolar interactions and
anisotropy. In case (c), the location of the canting transition
is independent on the strength of dipolar interactions
and anisotropy. We observe a maximum of the specific heat
at the canting transition in cases (b) and (c).
Another conclusion is that, due to the anisotropy
of the system, we predict an anisotropy in the specific heat.
Namely, the specific heat in a perpendicular magnetic field
is larger than the specific heat in an in--plane magnetic field.

However, there is also a contribution
to the specific heat due to non magnetic
interlayers. As only a few percentage
of their conduction electrons is
polarized by the exchange coupling
\cite{Staehler} the interlayers
should behave nearly as a bulk paramagnet.
This paramagnetic contribution per spin has then the form
\begin{equation}
c_v^{(para)} = \frac{(\beta h)^{2}}
{\cosh^{2}{(\beta h)}}
,
\end{equation}
where $\beta$ is the inverse temperature.
This paramagnetic contribution is much larger than the
spin wave contribution which is a source of
experimental difficulty.  Clearly, the sample should be
processed in such a way as to minimize the number of
paramagnetic atoms.

\subsection*{Acknowledgements}
We would like to thank Monique Giroud for useful discussions.
R.M. acknowledges discussions with M. Fabrizio.
R.M. also acknowledges the hospitality of NEC Research Institute
at Princeton and CRTBT at Grenoble
where part of this job was done.

\newpage
\renewcommand\textfraction{0}
\renewcommand
\floatpagefraction{0}
\noindent {\bf Figure captions}

\begin{figure}[h]
\caption{}
\label{Fig1}
Dispersion relation as a function of the magnetic field.
The dispersion relation is plotted along the line
$q_x=q_y=0$ as a function of the strength of the
magnetic field in the extended Brillouin zone
$[-\pi,\pi]$. There is no dipolar interaction
nor exchange interaction. We took $J_{\perp}=1$.
\end{figure}

\begin{figure}[h]
\caption{}
\label{Fig2}
Specific heat per spin in the different cases considered
in this article, with $J_{\perp}=1 K$, $J_{\parallel}=100 K$,
and as a function of the external magnetic field.
The curve (a) corresponds to the specific heat in the
absence of dipolar interactions and anisotropy.
(b) corresponds to a perpendicular applied magnetic field,
with $J_d = 1 K$ and $K = 0.5 K$.
(c) corresponds to a longitudinal magnetic field with
$J_d=1 K$ and $J_{\perp} = 0.5 K$.
\end{figure}

\begin{figure}[h]
\caption{}
\label{Fig3}
Dispersion relation as a function of
the intensity of a perpendicular magnetic field.
The dispersion relation is plotted along the line
$q_x=q_y=0$ as a function of the strength of the
magnetic field in the extended Brillouin zone
$[-\pi,\pi]$. We took $J_{\perp}=1$, $J_d=1$
and $K=0$. Notice the differences
with figure \ref{Fig1}.
\end{figure}

\begin{figure}[h]
\caption{}
\label{Fig4}
Dispersion relation as a function of the
intensity of a perpendicular magnetic field.
The dispersion relation is plotted along the line
$q_x=q_y=0$ as a function of the strength of the
magnetic field in the extended Brillouin zone
$[-\pi,\pi]$. We took $J_{\perp}=1$, $J_d=1$
and $K=0.5$. The variations of the
dispersion relation are very similar to
the ones of figure \ref{Fig3} where
$K$ was set to zero.
\end{figure}

\begin{figure}[h]
\caption{}
\label{Fig5}
Dispersion relation as a function of the
intensity of a longitudinal magnetic field.
The dispersion relation is plotted along the line
$q_x=q_y=0$ as a function of the strength of the
magnetic field in the extended Brillouin zone
$[-\pi,\pi]$. We took $J_{\perp}=1$, $J_d=1$
and $K=0.5$. Notice the differences
with figure \ref{Fig1}.
\end{figure}

\begin{thebibliography}{99}
\bibitem{remarque} As the thickness of the ferromagnetic
layers is increased, the strength of dipolar interactions
$J_d$ increases whereas the anisotropy $K$ remains
the same.
\end{thebibliography}
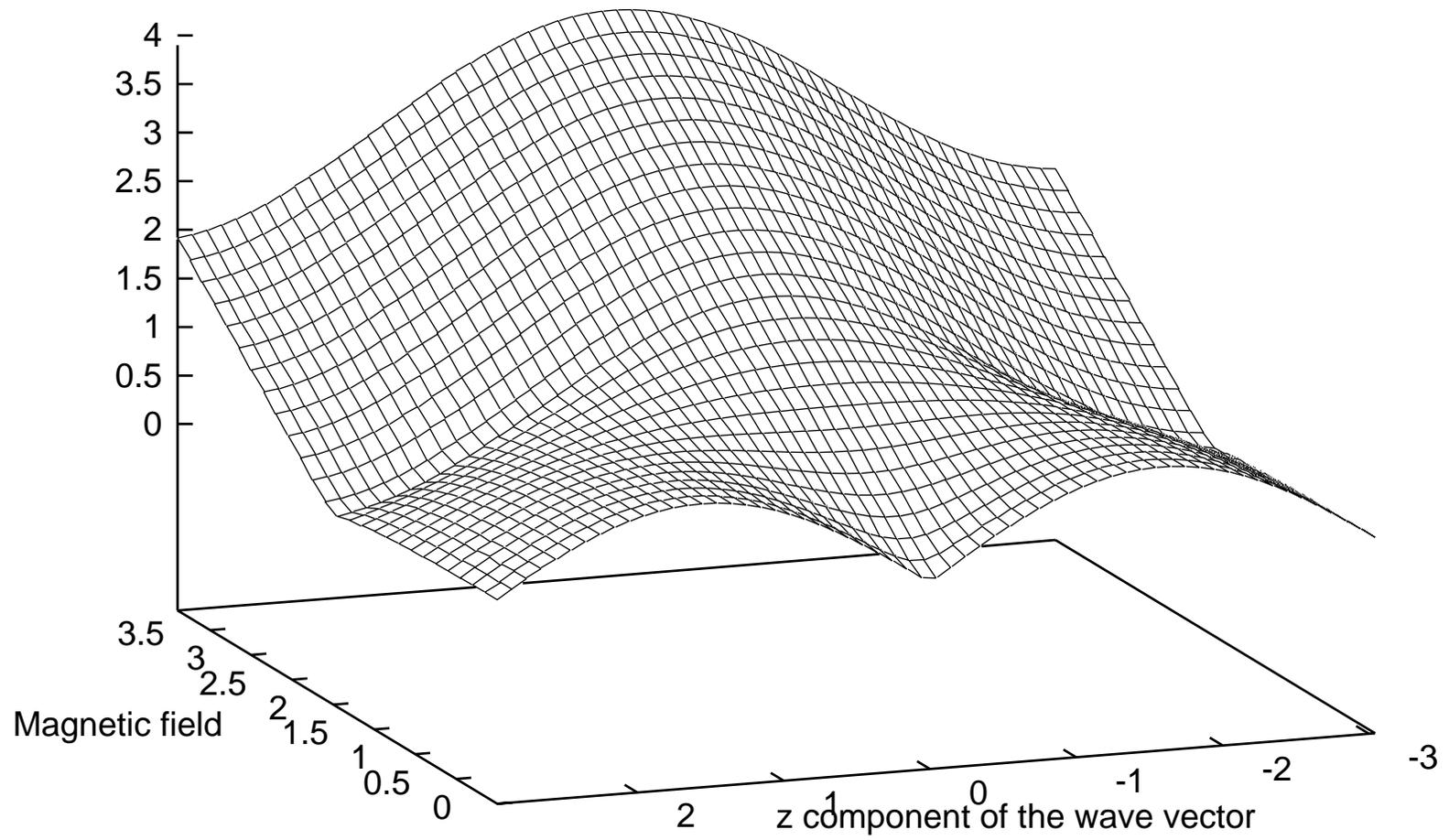
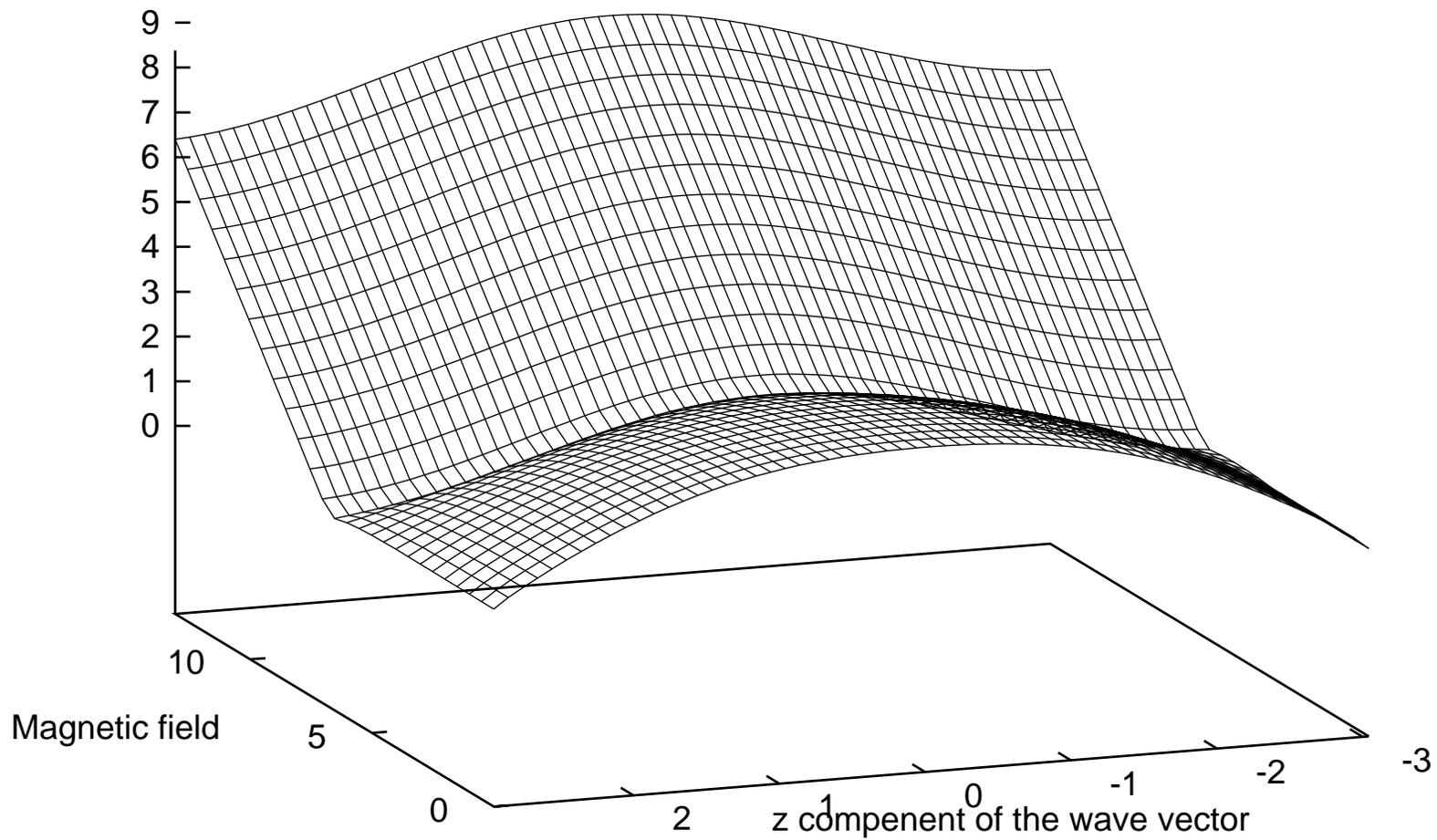
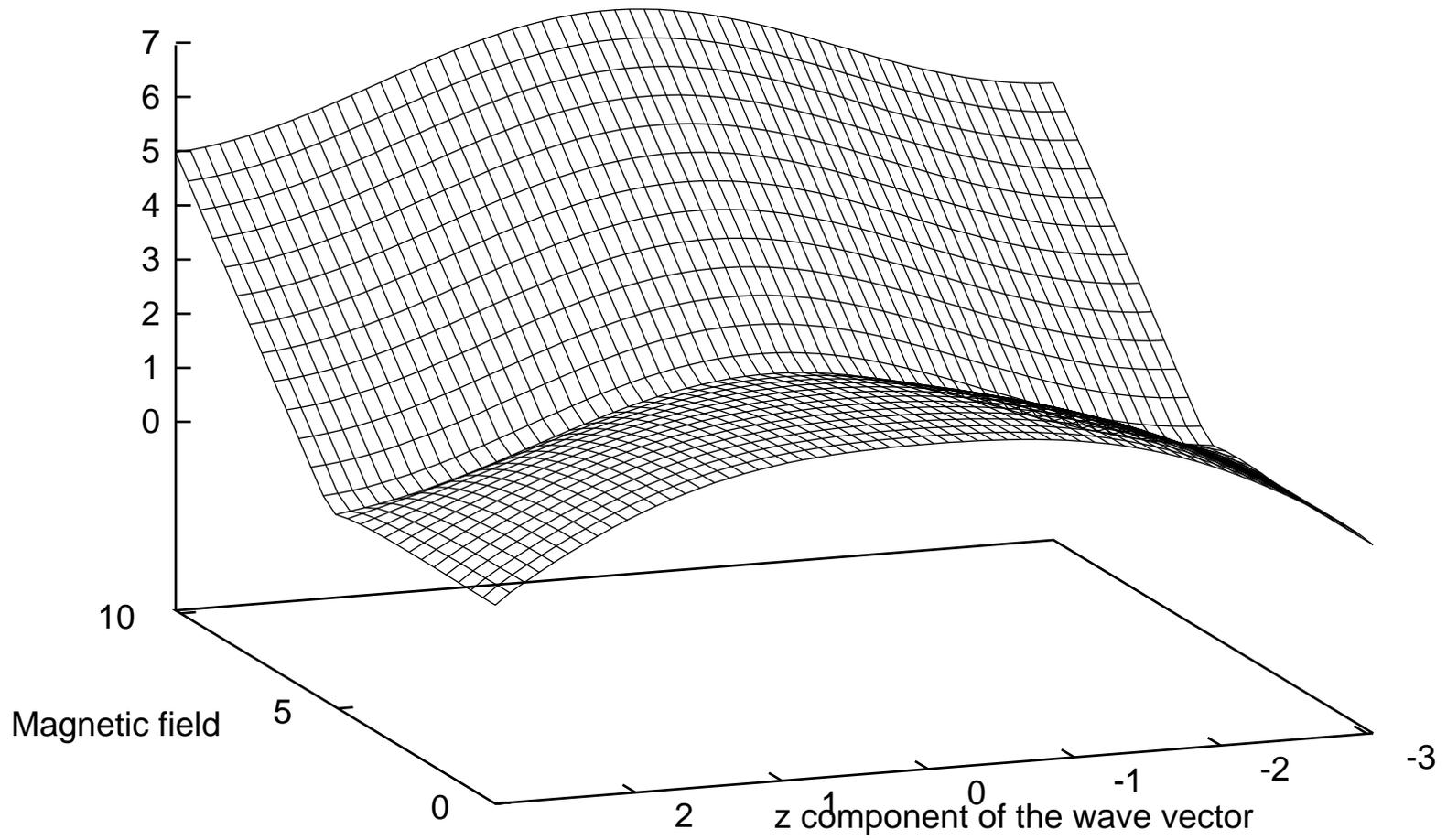
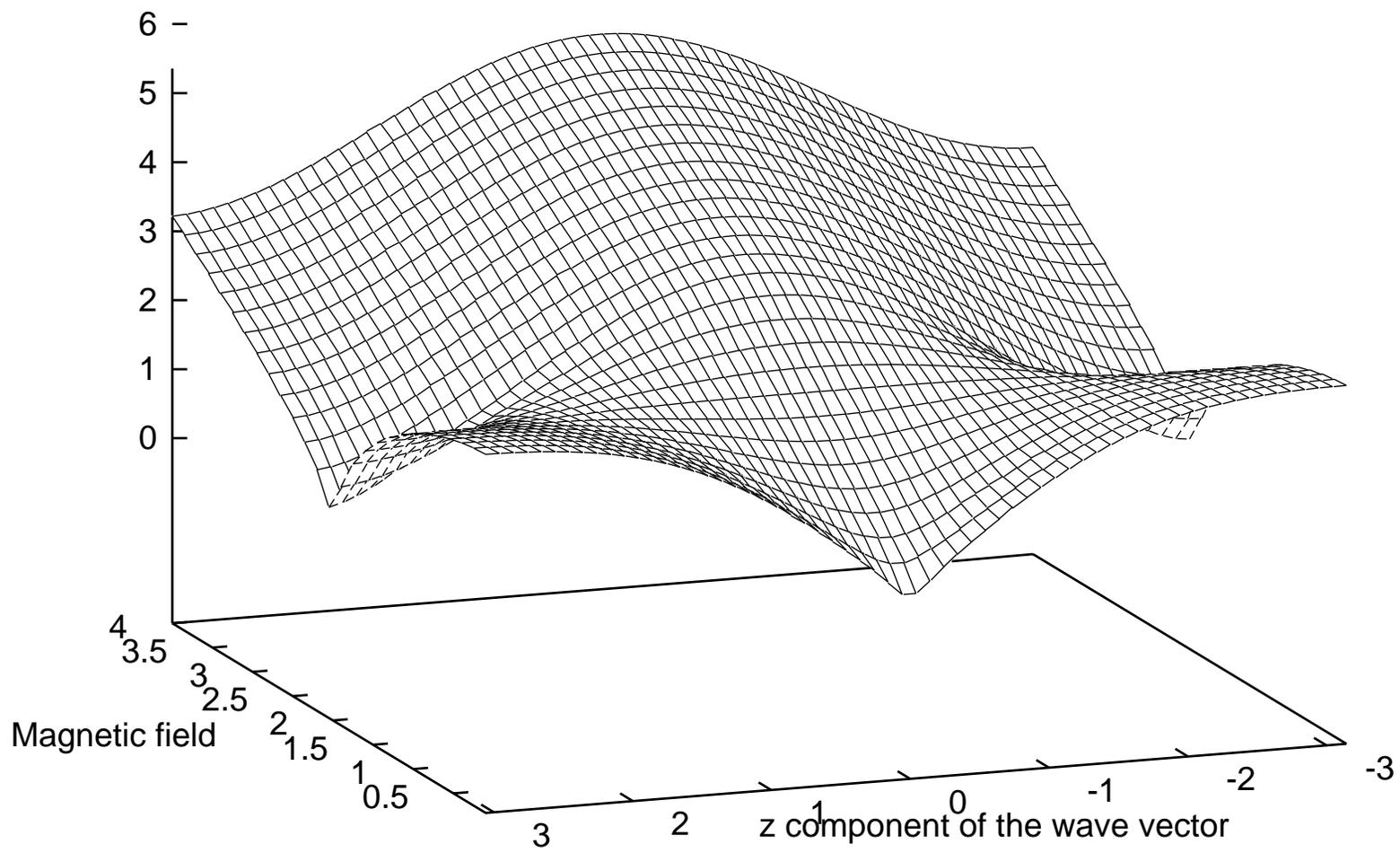
\end{document}